\documentclass[11pt]{article}

\usepackage[final]{acl}

\usepackage{times}
\usepackage{latexsym}

\usepackage[T1]{fontenc}

\usepackage[utf8]{inputenc}

\usepackage{microtype}

\usepackage{inconsolata}

\usepackage{graphicx}

\usepackage{graphicx} 
\usepackage{natbib}  
\usepackage{caption} 
\usepackage{algorithm}
\usepackage{algorithmic}

\usepackage{newfloat}
\usepackage{listings}
\DeclareCaptionStyle{ruled}{labelfont=normalfont,labelsep=colon,strut=off} 
\floatstyle{ruled}
\newfloat{listing}{tb}{lst}{}
\floatname{listing}{Listing}

\usepackage{booktabs}
\usepackage{multirow}

\usepackage[most]{tcolorbox}
\usepackage{enumitem}
\usepackage{amssymb}
\usepackage{svg}

\usepackage{amsmath}
\usepackage{array}
\title{RecHarness: A Bandit-Routed Agentic Harness for Self-Evolving Recommender Systems}

\author{
  \textbf{Haoran Ling\textsuperscript{1,2}}
  \thanks{These authors contributed equally.}
  \thanks{Work done during an internship at Kuaishou Technology.},
  \textbf{Yuecheng Li\textsuperscript{3,*}},
  \textbf{Zeyu Song\textsuperscript{3,*}},
  \textbf{Jing Yao\textsuperscript{3}}
  \\
  \textbf{Shuwen Kang\textsuperscript{4}},
  \textbf{Songcheng Ren\textsuperscript{4}},
  \textbf{Chi Lu\textsuperscript{3}}
  \thanks{Corresponding author.},
  \textbf{Wenjin Wu\textsuperscript{3}},
  \textbf{Peng Jiang\textsuperscript{3}}
  \\
  \textsuperscript{1}Georgia Institute of Technology,
  \textsuperscript{2}Tianjin University
  \\
  \textsuperscript{3}Kuaishou Technology, China,
  \textsuperscript{4}Peking University
  \\
  \texttt{hling34@gatech.edu};
  \texttt{\{liyuecheng, songzeyu, luchi, wuwenjin, jiangpeng\}@kuaishou.com}
}

\begin{document}
\maketitle
\begin{abstract}
Optimizing recommender models remains heavily reliant on manual iteration over architectures, objectives, and training strategies. While LLM-based agents can automate this trial-and-error process, relying on LLMs to both select modification directions and formulate concrete hypotheses can lead to unstable search under limited trial budgets. Inspired by the above challenge, we propose \textbf{RecHarness}, a \textit{Bandit-Routed Agentic Harness} for automated recommender model optimization. RecHarness decouples the optimization process into two steps: a bandit router selects the next modification direction according to historical validation feedback, while the LLM generates a concrete optimization hypothesis and executable code edit within the selected direction. To sustain long-horizon exploration, RecHarness uses a jump-basin mechanism to activate a structural-jump arm when local edits stagnate. Across multiple recommendation tasks, datasets, and model backbones, RecHarness achieves more consistent performance improvements and uses limited trial budgets more effectively than LLM-directed search. During a 7-day online A/B test on a large-scale short-video advertising platform, the selected candidate improves ADVV by 2.084\%, Revenue by 0.534\%, and Exposure by 0.559\%. Code is available at: \url{https://github.com/6lyc/RecHarness}.
\end{abstract}

\section{Introduction}

LLM-based machine learning engineering (MLE) agents have made substantial progress on automated model development. Given a dataset, an evaluation metric, and an initial codebase, these agents can inspect task descriptions, write or modify training code, execute experiments, and iteratively improve candidate solutions using validation results and execution logs \cite{huang2023mlagentbench,chan2025mlebench,jiang2025aide,nam2025mlestar}. Existing methods often formulate this process as search over candidate programs, using multi-branch search, tree search, or graph search to explore possible solutions and validation scores to select promising branches. Recent work such as Reasoning-as-Gradient \cite{zhang2026reasoningasgrad} argues that execution feedback should not be compressed only into scalar scores: error logs, training dynamics, and validation outcomes can also serve as textual optimization signals that guide more directed code updates \cite{pryzant2023automaticprompt,shinn2023reflexion,madaan2023selfrefine,yuksekgonul2024textgrad,zhang2026reasoningasgrad}. This line of work suggests that LLM reasoning is valuable not only for generating code, but also for interpreting experimental feedback and forming the next improvement hypothesis.

Recommender systems (RecSys) optimization is a natural and practically important setting for LLM-based MLE agents. Unlike one-shot model construction, recommender development is often a continuous training-code iteration process. In sequential recommendation, ranking and click prediction, and watch-time prediction, model quality is affected by the backbone architecture, training objective, optimization strategy, regularization, feature interaction design, sequence modeling choice, and implementation details. Therefore, the optimization target is not a single fixed recommender template, but a broader engineering search space whose best solution depends on the dataset, task formulation, model family, and evaluation objective. Recent agentic recommender work also emphasizes that modern recommender optimization still relies heavily on iterative engineering, and that agent-based systems can help automate this loop by using persistent memory and evolving optimization methodology \cite{cheng2026agentsteer,ou2026deepresearch,mu2026evorec}. \textit{Our goal is therefore not to optimize a single recommender template, but to build a validation-guided harness that can adapt to different recommendation scenarios, model families, and evaluation objectives.}

However, directly applying a general-purpose MLE agent to recommender optimization leaves an important systems challenge. Each candidate modification typically requires code generation, execution, training, and validation, so every trial consumes non-negligible budget. Moreover, Recommender model optimization is not a set of independent one-shot experiments, but a continuous process of model iteration. Each trial provides optimization-relevant evidence beyond its final validation score, such as convergence behavior, training stability, and exposed failure modes or bottlenecks. These signals must be interpreted in the context of the current model state, training dynamics, and previous modifications. Therefore, an automated optimizer cannot rely solely on scalar validation feedback; it must interpret logs, trends, and failure signals from each round, while accumulating cross-round evidence to determine which optimization directions are effective. Recent studies on LLM-based sequential decision making provide a useful motivation for this design. In-context bandit experiments show that general-purpose LLMs may require external summaries or algorithmic support to explore reliably \cite{krishnamurthy2024llmexplore}, while LLM-enhanced multi-armed bandit methods suggest that combining LLM reasoning or prediction with classical bandit mechanisms can be more effective than direct LLM arm selection \cite{sun2026llmenhancedmab}. \textbf{These findings align with our setting: validation feedback should be used both as textual feedback for LLM reasoning and as an explicit posterior state for guiding future trials.}

Based on the above findings, we introduce \textbf{RecHarness}, a validation-guided optimization harness for self-evolving recommender systems. RecHarness organizes optimization as a sequence of isolated training-validation trials. At each round, it maintains candidate optimization arms defined for the target recommender setting, updates their posterior state using validation feedback, and uses Thompson-style routing to select promising arms under a limited trial budget. Given the selected arms, the LLM uses the current incumbent, execution logs, validation trends, and a dynamically updated Experiment Skill distilled from experiment memory to form concrete improvement hypotheses and generate executable code modifications. For high-impact changes such as architecture, loss changes, RecHarness further supports a retuning window so that the system can evaluate whether a structural jump becomes beneficial after local adaptation. \textbf{RecHarness therefore combines LLM reasoning with validation-driven posterior routing: scalar validation evidence accumulates across trials to decide which optimization directions to try next, while textual feedback helps the LLM generate concrete hypotheses and executable code mutations within the selected directions.}

Our contributions are:
\begin{itemize}
    \item We introduce RecHarness, the first optimization harness for self-evolving recommender systems, designed to support diverse recommendation scenarios.

    \item We design a bandit-routed optimization mechanism that jointly leverages scalar scores and textual feedback through bandit routing and LLM reasoning.

    \item We conduct empirical studies across recommender tasks, datasets, and model templates, demonstrating the effectiveness of RecHarness in improving recommendation performance under limited trials, and an online A/B test confirms its gains in business scenarios.
\end{itemize}

\begin{figure*}[t]
\centering
\includegraphics[width=0.92\textwidth]{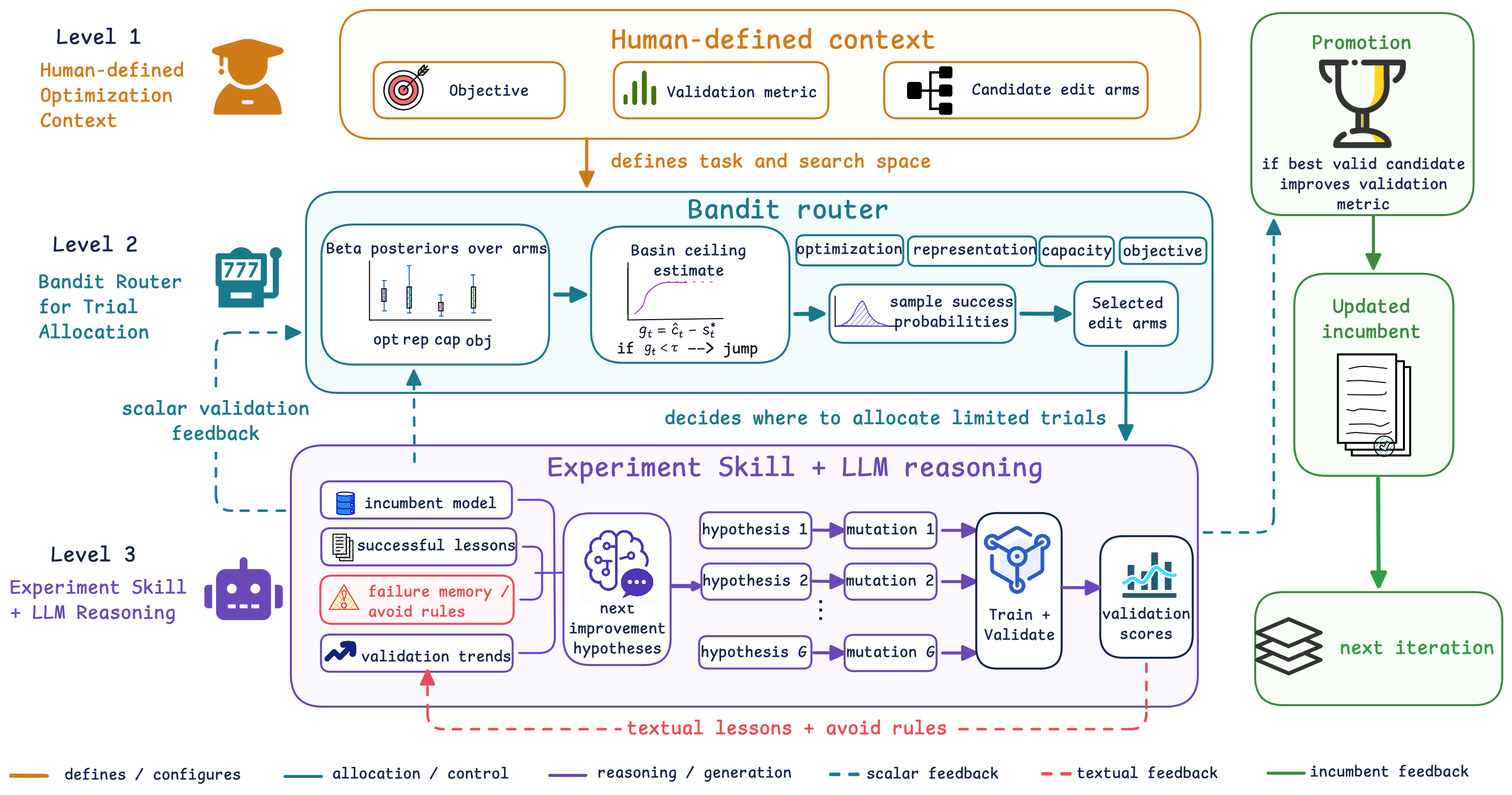}
\caption{Overview of RecHarness. RecHarness operates across three levels: \textbf{Level 1} defines the human-specified optimization context, including the objective, validation metric, and candidate edit arms; \textbf{Level 2} performs bandit-based trial allocation using scalar validation feedback; and \textbf{Level 3} applies Experiment Skill and LLM reasoning to generate hypotheses, executable mutations, validation scores, and textual lessons. Validated improvements are promoted as the updated incumbent for the next iteration.}
\label{fig:recharness-overview}
\end{figure*}

\section{Methodology}

RecHarness organizes recommender optimization through following complementary mechanisms: a bandit router allocates trials across human-defined edit arms, Experiment Skill guides LLM-generated mutations using accumulated evidence, and a basin-aware jump mechanism enables structural exploration after local progress stalls. Human experts specify the objective, validation metric, and candidate arms. Figure~\ref{fig:recharness-overview} provides an overview.

\subsection{Problem Formulation and Search Loop}

Given an initial recommender implementation $f_0$, a training-validation split $\mathcal{D}$, a validation metric $M$, and a total budget $B$, RecHarness seeks the best validated implementation through sequential code mutations. Each trial generates, trains, and evaluates a candidate:
\begin{equation}
\begin{aligned}
    & \omega_f = \operatorname{Train}(f,\mathcal{D}_{\mathrm{train}}),\\
    & M(f) = M(\omega_f;\mathcal{D}_{\mathrm{val}}),\\
    & \sum_t\sum_{a\in\mathcal{G}_t}c(f_{t,a}) \le B,
\end{aligned}
\end{equation}
where $\mathcal{G}_t$ is the selected trial group and $c(f_{t,a})$ includes generation, execution, training, and validation costs. For sequential recommendation, we use validation HR@10.

The search space is organized into predefined edit arms $\mathcal{A}=\{a_1,\ldots,a_K\}$. Each arm specifies an interpretable modification direction, such as regularization, architecture, pooling, features, or loss design, rather than an exact patch.

Let $f_t^\star$ be the incumbent before round $t$, with score $s_t^\star=M(f_t^\star)$. For each selected arm, a frozen LLM generates a candidate conditioned on the incumbent, Experiment Skill $m_t$, and recent logs $\ell_t$:
\begin{equation}
\begin{aligned}
    f_{t,a} &= \mu_\phi(f_t^\star,a,m_t,\ell_t),\\
    f_{t+1}^\star
    &\in \arg\max_{f\in\{f_t^\star\}\cup\mathcal{V}_t}M(f),
\end{aligned}
\end{equation}
where $\mathcal{V}_t$ contains valid candidates eligible for promotion. The incumbent is retained unless a candidate strictly improves its score. Ordinary local candidates are evaluated directly; jump candidates become eligible only after the retuning procedure described below.

\subsection{Thompson Sampling for Trial Allocation}

RecHarness uses Thompson sampling to allocate trials across edit arms. Each arm maintains a Beta posterior, and each round selects the $G$ largest posterior samples among currently available arms:
\begin{equation}
\begin{aligned}
    \tilde{\theta}_a
    &\sim \operatorname{Beta}(\alpha_a,\beta_a),\\
    \mathcal{G}_t
    &= \operatorname{TopG}_{a\in\mathcal{A}_t}
    (\tilde{\theta}_a).
\end{aligned}
\end{equation}
The selected arms define semantic search directions, while the LLM determines their concrete implementations. Grouped candidates are executed in parallel from the same incumbent.

For a valid local trial, define its improvement as $\Delta_t(a)=M(f_{t,a})-s_t^\star$, and let $\bar{\Delta}_t$ be the mean improvement among valid candidates in the group. A trial succeeds only if it exceeds the group mean and does not underperform the incumbent:
\begin{equation} \small
    r_t(a)=
    \begin{cases}
        \mathbb{I}\!\left[
            \Delta_t(a)>\bar{\Delta}_t
            \land \Delta_t(a)\ge0
        \right],
        & f_{t,a}\text{ is valid},\\
        0, & \text{otherwise}.
    \end{cases}
\end{equation}
This criterion is equivalent to requiring a positive group-standardized advantage together with incumbent non-degradation. The corresponding posterior update is
\begin{equation}
\begin{aligned}
    \alpha_a &\leftarrow \alpha_a+r_t(a),\\
    \beta_a &\leftarrow \beta_a+1-r_t(a).
\end{aligned}
\end{equation}

Because outcomes depend on the incumbent and previous edits, RecHarness does not assume a fixed global success probability for each arm. Posterior statistics instead provide empirical evidence for balancing promising and uncertain directions under a limited budget. This explicit routing also separates arm allocation from LLM reasoning, following evidence that direct LLM arm selection can be unreliable \cite{krishnamurthy2024llmexplore,sun2026llmenhancedmab}.

\subsection{Experiment Skill for Evidence-Guided Mutation}

After arm selection, Experiment Skill guides the LLM in forming a concrete improvement hypothesis. It is a compact textual state containing the current incumbent, successful edits, rejected directions, failure reasons, and recent validation trends.

After each round, RecHarness distills successful trials into reusable lessons and failed or harmful trials into avoid rules, then refreshes the Skill using trial summaries:
\begin{equation} \small
    m_{t+1}
    =\operatorname{Summarize}\!\left(
        m_t,
        \{a,f_{t,a},M(f_{t,a}),e_{t,a}\}_{a\in\mathcal{G}_t}
    \right),
\end{equation}
where $e_{t,a}$ records execution status, errors, and compact training logs; validation scores are included when available.

The two feedback channels have distinct roles: scalar outcomes update the bandit posterior to determine where to search, while textual feedback conditions $\mu_\phi$ to determine how to modify the model within a selected arm. Experiment Skill does not rank arms or modify posterior updates.

\subsection{Basin-Aware Jump and Retuning}

Incumbent-based refinement may saturate within a local basin. RecHarness therefore partitions the arms into disjoint local and jump sets, $\mathcal{A}=\mathcal{A}_{\mathrm{local}}\cup\mathcal{A}_{\mathrm{jump}}$. Local arms make incremental changes, whereas jump arms introduce higher-level structural changes.

Jump arms become available when the incumbent's improvement rate over the preceding $W$ rounds falls below a threshold $\tau$:
\begin{equation}
\begin{aligned}
    \widehat{v}_t
    &= \frac{s_t^\star-s_{t-W}^\star}{W},\\
    \mathcal{A}_t
    &=
    \begin{cases}
        \mathcal{A}_{\mathrm{local}}\cup
        \mathcal{A}_{\mathrm{jump}},
        & \widehat{v}_t\le\tau,\\
        \mathcal{A}_{\mathrm{local}},
        & \text{otherwise}.
    \end{cases}
\end{aligned}
\end{equation}
Eligible jump arms compete via Thompson sampling; if selected, the highest-sampled jump executes alone.

A selected jump candidate initializes a temporary branch with its own local incumbent and receives up to $R$ local retuning rounds. Let $f_{t,r}^{(a)}$ denote the branch's best implementation after $r$ rounds, with $r=0$ representing the initial jump candidate. The branch is accepted only if its best retuned score exceeds the pre-jump incumbent by a margin $\delta_{\mathrm{jump}}>0$:
\begin{equation} \small
    \operatorname{AcceptJump}(a,t)
    =\mathbb{I}\!\left[
        \max_{0\le r\le R}M(f_{t,r}^{(a)})
        -s_t^\star>\delta_{\mathrm{jump}}
    \right].
\end{equation}
Upon acceptance, the best branch implementation becomes eligible for global promotion; otherwise, the branch is discarded. All retuning trials count toward the total budget. This delayed evaluation allows structural changes to benefit from local adaptation before acceptance.

\section{Experiments}
\begin{table*}[t]
\centering
\small
\renewcommand{\arraystretch}{0.88}
\setlength{\aboverulesep}{0.3ex}
\setlength{\belowrulesep}{0.3ex}
\setlength{\tabcolsep}{1.6mm}
\begin{tabular}{@{}ll*{10}{c}@{}}
\toprule
\multirow{2}{*}{Dataset}
& \multirow{2}{*}{Method}
& \multicolumn{2}{c}{GRU4Rec}
& \multicolumn{2}{c}{BERT4Rec}
& \multicolumn{2}{c}{NextItNet}
& \multicolumn{2}{c}{SASRec}
& \multicolumn{2}{c}{HSTU} \\
\cmidrule(lr){3-4}
\cmidrule(lr){5-6}
\cmidrule(lr){7-8}
\cmidrule(lr){9-10}
\cmidrule(lr){11-12}
& & N@10 & H@10
& N@10 & H@10
& N@10 & H@10
& N@10 & H@10
& N@10 & H@10 \\
\midrule

\multirow{4}{*}{Movies}
& Base
& 0.1267 & 0.2317
& 0.2585 & 0.4302
& 0.1347 & 0.2617
& 0.3688 & 0.5335
& 0.3253 & 0.5038 \\
& MLE-STAR
& 0.3384 & 0.5331
& 0.3711 & 0.5677
& 0.3282 & 0.5072
& 0.3891 & 0.5667
& 0.4006 & 0.5814 \\
& Self-EvolveRec
& 0.3356 & 0.5188
& 0.3587 & 0.5579
& 0.3407 & 0.5219
& 0.3448 & 0.5330
& 0.3683 & 0.5472 \\
& \textbf{RecHarness}
& \textbf{0.3704} & \textbf{0.5560}
& \textbf{0.3883} & \textbf{0.5775}
& \textbf{0.3489} & \textbf{0.5353}
& \textbf{0.4023} & \textbf{0.5923}
& \textbf{0.4072} & \textbf{0.5917} \\
\midrule

\multirow{4}{*}{Scientific}
& Base
& 0.1663 & 0.2831
& 0.2379 & 0.4081
& 0.2179 & 0.3632
& 0.2805 & 0.4701
& 0.2804 & 0.4690 \\
& MLE-STAR
& 0.3038 & 0.5080
& 0.3189 & 0.5221
& 0.3115 & 0.5076
& 0.3217 & 0.5069
& 0.3431 & 0.5418 \\
& Self-EvolveRec
& 0.2922 & 0.4943
& 0.3102 & 0.5213
& 0.3160 & 0.5160
& 0.2898 & 0.4819
& 0.3049 & 0.4857 \\
& \textbf{RecHarness}
& \textbf{0.3324} & \textbf{0.5301}
& \textbf{0.3281} & \textbf{0.5328}
& \textbf{0.3216} & \textbf{0.5373}
& \textbf{0.3311} & \textbf{0.5381}
& \textbf{0.3516} & \textbf{0.5503} \\
\midrule

\multirow{4}{*}{Electronics}
& Base
& 0.1789 & 0.3005
& 0.1870 & 0.3186
& 0.2293 & 0.3833
& 0.2614 & 0.4271
& 0.2424 & 0.4038 \\
& MLE-STAR
& 0.2377 & 0.4007
& 0.2555 & 0.4199
& 0.2583 & 0.4179
& 0.2594 & 0.4185
& 0.2570 & 0.4247 \\
& Self-EvolveRec
& 0.2353 & 0.3932
& 0.2381 & 0.3950
& 0.2557 & 0.4227
& 0.2626 & 0.4262
& 0.2279 & 0.3801 \\
& \textbf{RecHarness}
& \textbf{0.2454} & \textbf{0.4011}
& \textbf{0.2742} & \textbf{0.4426}
& \textbf{0.2686} & \textbf{0.4302}
& \textbf{0.2635} & \textbf{0.4336}
& \textbf{0.2671} & \textbf{0.4291} \\
\midrule

\multirow{4}{*}{CDs}
& Base
& 0.1394 & 0.2586
& 0.2872 & 0.4693
& 0.1367 & 0.2544
& 0.2614 & 0.5833
& 0.3192 & 0.5124 \\
& MLE-STAR
& 0.3536 & 0.5577
& 0.3840 & 0.5966
& 0.3479 & 0.5498
& 0.3961 & 0.5882
& 0.4021 & 0.6014 \\
& Self-EvolveRec
& 0.3088 & 0.5087
& 0.3730 & 0.5975
& 0.3428 & 0.5439
& 0.2978 & 0.4872
& 0.4100 & 0.5966 \\
& \textbf{RecHarness}
& \textbf{0.3925} & \textbf{0.6034}
& \textbf{0.4240} & \textbf{0.6276}
& \textbf{0.3516} & \textbf{0.5516}
& \textbf{0.4465} & \textbf{0.6593}
& \textbf{0.4481} & \textbf{0.6517} \\
\midrule

\multirow{4}{*}{Average}
& Base
& 0.1528 & 0.2685
& 0.2427 & 0.4066
& 0.1797 & 0.3157
& 0.2930 & 0.5035
& 0.2918 & 0.4723 \\
& MLE-STAR
& 0.3084 & 0.4999
& 0.3324 & 0.5266
& 0.3115 & 0.4956
& 0.3416 & 0.5201
& 0.3507 & 0.5373 \\
& Self-EvolveRec
& 0.2930 & 0.4788
& 0.3200 & 0.5179
& 0.3138 & 0.5011
& 0.2988 & 0.4821
& 0.3278 & 0.5024 \\
& \textbf{RecHarness}
& \textbf{0.3352} & \textbf{0.5227}
& \textbf{0.3537} & \textbf{0.5451}
& \textbf{0.3227} & \textbf{0.5136}
& \textbf{0.3609} & \textbf{0.5558}
& \textbf{0.3685} & \textbf{0.5557} \\
\bottomrule
\end{tabular}
\caption{Performance on Amazon Reviews, averaged over three runs. N@10 and H@10 denote NDCG@10 and HR@10. Average reports the arithmetic mean over the four datasets. \textbf{Bold} indicates the best result for each model}
\label{tab:amazon-overall}
\end{table*}

\subsection{Experimental Settings}

\textbf{Tasks and datasets.}
Amazon Reviews~\cite{hou2026bridging} contains user--item review interactions; we use its Movies, Scientific, Electronics, and CDs subsets with five-core filtering for next-item prediction.
KuaiRec~\cite{gao2022kuairec} is a dense user--video interaction dataset used for watch-time, watch-ratio, and ranking prediction.
We report NDCG@10 and HR@10 on Amazon, and XAUC and MAE on KuaiRec.
Dataset details are provided in Appendix~\ref{app:data}.

\textbf{Baselines.}
We compare RecHarness with MLE-STAR~\cite{nam2025mlestar}, and Self-EvolveRec~\cite{kim2026self}.
Details are provided in Appendix~\ref{app:base}.

\textbf{Optimized recommendation models.}
On Amazon, we optimize GRU4Rec~\cite{hidasi2016gru4rec}, BERT4Rec~\cite{sun2019bert4rec}, NextItNet~\cite{yuan2019simple}, SASRec~\cite{kang2018sasrec}, and HSTU~\cite{zhai2024actions}.
On KuaiRec, we optimize D2Q~\cite{zhan2022deconfounding}, TPM~\cite{lin2023tree}, and GR~\cite{ma2024generative}, covering diverse architectures and feedback objectives. Details are provided in Appendix~\ref{app:templates}.

\textbf{Search protocol.}
Each run starts from an initial template, evaluates the baseline, and iteratively searches over candidate edit arms.
Candidate selection, routing updates, and version promotion use validation feedback only.
The test set is evaluated once after search.

\textbf{Trial execution and budget.}
Candidates run in parallel in isolated workspaces, returning validation scores, logs, and execution status.
Only executable candidates that outperform the incumbent are promoted.
Each model receives a GPU-time budget of 43,200 seconds; search stops when the remaining budget cannot support another trial group.

\subsection{Overall Performance}


RecHarness achieves a favorable balance between recommendation
performance and LLM token usage.
On Amazon, it obtains the highest average NDCG@10 and HR@10
across all five backbones (Table~\ref{tab:amazon-overall}).
On SASRec, RecHarness leads on both metrics across all four
datasets, improving average NDCG@10 and HR@10 over MLE-STAR
by 5.64\% and 6.87\%, and over Self-EvolveRec by 20.79\%
and 15.30\%, respectively.
On KuaiRec, it achieves the best results on all 12
model--metric combinations, demonstrating optimization
benefits across architectures and tasks
(Table~\ref{tab:kuairec-overall}).

Meanwhile, the three recorded SASRec runs show that
RecHarness reduces average total token usage by 31.99\%
relative to MLE-STAR and 96.94\% relative to Self-EvolveRec
(Table~\ref{tab:sasrec-llm-usage}).
Despite comparable call counts to MLE-STAR, it uses 90.54\%
fewer output tokens.
This advantage may stem from its explicit division of roles:
the router allocates search directions using validation
feedback, structured edit arms constrain the reasoning scope,
and Experiment Skill guides concrete mutations through
compact accumulated experience.
Together, these mechanisms can improve search efficiency
and reduce repeated exploration.

\subsection{Ablation Study}

\begin{table}[t]
\centering
\small
\setlength{\tabcolsep}{3pt}
\resizebox{\columnwidth}{!}{%
\begin{tabular}{@{}llcccc@{}}
\toprule
Model & Method
& WT-XAUC $\uparrow$
& WT-MAE $\downarrow$
& WR-XAUC $\uparrow$
& WR-MAE $\downarrow$ \\
\midrule
\multirow{4}{*}{D2Q}
& Base           & 0.5310 & 3.4129 & 0.7307 & 0.3558 \\
& MLE-STAR       & 0.5635 & 3.3657 & 0.7375 & 0.3509 \\
& Self-EvolveRec & 0.5522 & 3.3890 & 0.7340 & 0.3533 \\
& \textbf{RecHarness}
& \textbf{0.5947} & \textbf{3.3107}
& \textbf{0.7444} & \textbf{0.3451} \\
\midrule
\multirow{4}{*}{TPM}
& Base           & 0.5342 & 4.5372 & 0.7012 & 0.4730 \\
& MLE-STAR       & 0.5722 & 4.1245 & 0.7142 & 0.4239 \\
& Self-EvolveRec & 0.5405 & 4.3210 & 0.7079 & 0.4401 \\
& \textbf{RecHarness}
& \textbf{0.5808} & \textbf{3.3406}
& \textbf{0.7418} & \textbf{0.3481} \\
\midrule
\multirow{4}{*}{GR}
& Base           & 0.6161 & 3.1901 & 0.7528 & 0.3323 \\
& MLE-STAR       & 0.6163 & 3.1980 & 0.7528 & 0.3331 \\
& Self-EvolveRec & 0.6163 & 3.1946 & 0.7529 & 0.3329 \\
& \textbf{RecHarness}
& \textbf{0.6176} & \textbf{3.1851}
& \textbf{0.7535} & \textbf{0.3319} \\
\bottomrule
\end{tabular}%
}
\caption{Performance on KuaiRec, averaged over three runs. WT and WR denote watch time and watch ratio; WT-MAE is measured in seconds.}
\label{tab:kuairec-overall}
\end{table}

\begin{table}[t]
\centering
\small
\setlength{\tabcolsep}{3pt}
\resizebox{\columnwidth}{!}{%
\begin{tabular}{@{}lrrrr@{}}
\toprule
Method
& Input tokens $\downarrow$
& Output tokens $\downarrow$
& Total tokens $\downarrow$
& LLM calls $\downarrow$ \\
\midrule
MLE-STAR
& 786,995
& 143,494
& 930,490
& \textbf{31.33} \\
Self-EvolveRec
& 20,049,255
& 644,409
& 20,693,664
& 613.00 \\
RecHarness
& \textbf{619,260}
& \textbf{13,573}
& \textbf{632,834}
& 31.67 \\
\bottomrule
\end{tabular}%
}
\caption{Recorded LLM usage on SASRec, averaged over three
runs per method.}
\label{tab:sasrec-llm-usage}
\end{table}

We ablate RecHarness on four Amazon datasets using SASRec, keeping the validation-driven update rule, GPU budget, and four parallel trials per round fixed.
The full method combines Thompson Routing (TR), Experiment Skill, and basin-aware jump and retuning.
Five variants replace TR with uniform random selection (\textbf{TR w/ Random}) or LLM selection from textual history (\textbf{TR w/ LLM}), remove predefined arms (\textbf{w/o Bandit}), remove Experiment Skill (\textbf{w/o Experiment Skill}), or disable jump and retuning (\textbf{w/o Jump}).

\begin{figure}[t]
\centering
\includegraphics[width=\columnwidth]{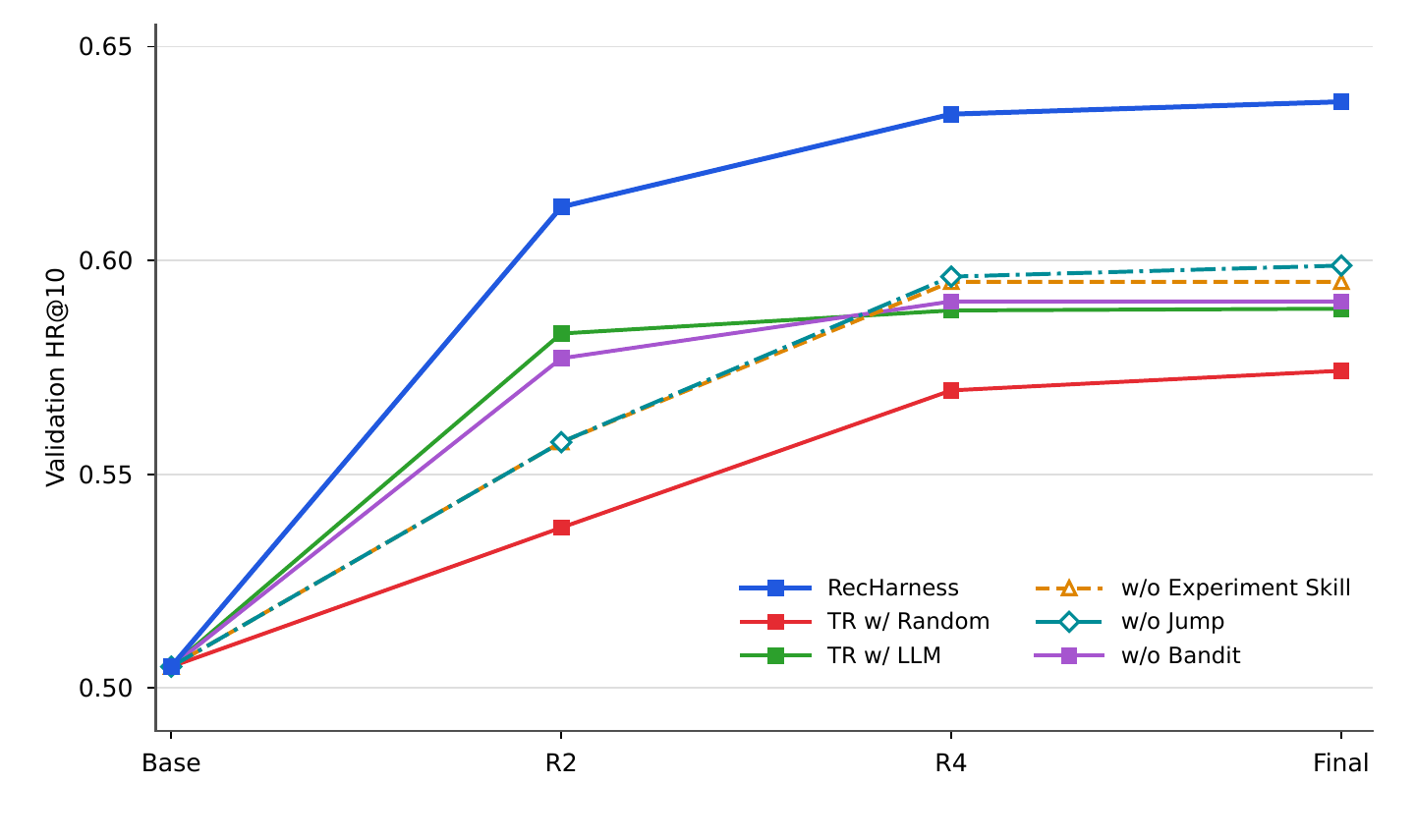}
\caption{Best-so-far validation HR@10 during SASRec search, macro-averaged over four Amazon datasets.}
\label{fig:rq2-progress}
\end{figure}

Figure~\ref{fig:rq2-progress} reports best-so-far validation HR@10 averaged over the four datasets.
RecHarness reaches 0.6125 by Round 2, exceeding all ablations' final scores, and finishes at 0.6371.
LLM routing improves over random routing (0.5887 vs.\ 0.5742) but trails posterior routing, while unrestricted editing reaches only 0.5904.
Removing Skill or Jump yields 0.5950 or 0.5988, respectively, with little progress after Round 4, suggesting that both components help sustain effective search.

\begin{figure}[t]
  \centering
  \begin{minipage}{0.90\columnwidth}
    \centering
    \includegraphics[width=\linewidth]{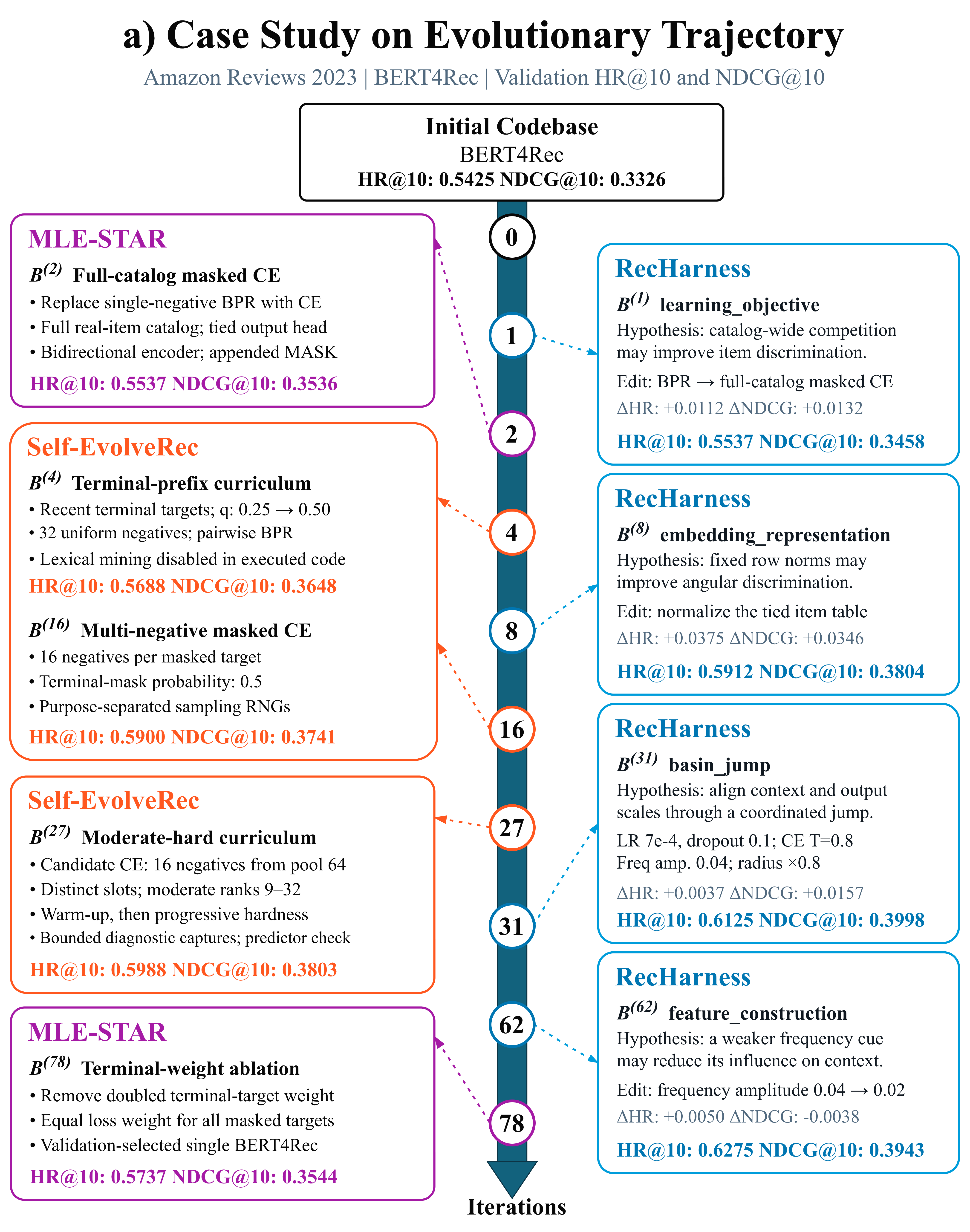}
    \par\vspace{2pt}
    \includegraphics[width=\linewidth]{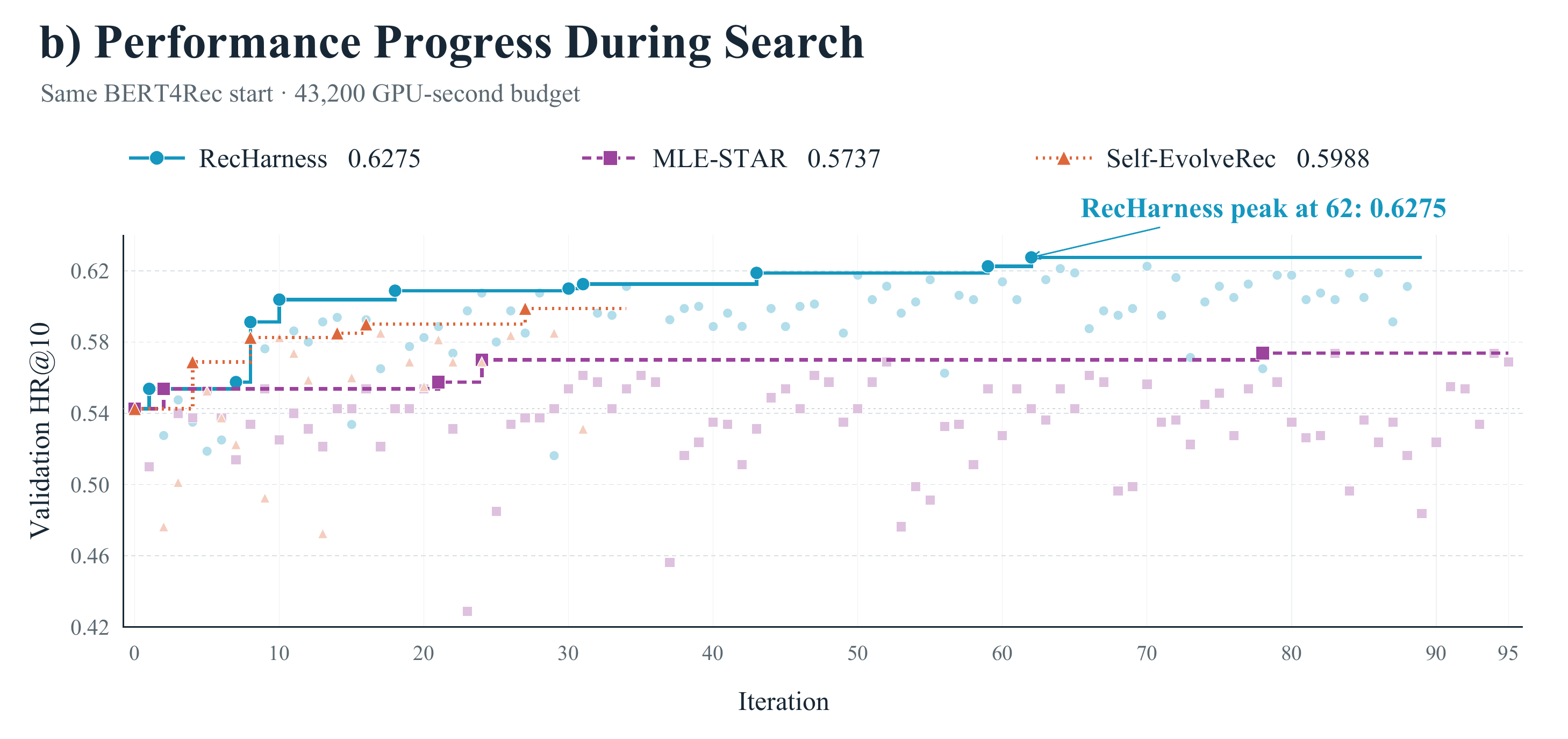}
  \end{minipage}
    \caption{Evolutionary trajectories across four Amazon categories
    (Seed Recommender: BERT4Rec).
    (a) Comparison of evolutionary paths.
    (b) Performance comparison across iterations.}
  \label{fig:case-study}
\end{figure}

\subsection{Case Study}

We compare RecHarness, MLE-STAR, and Self-EvolveRec using the same
BERT4Rec initialization and a 43,200 GPU-second budget cap.
Figure~\ref{fig:case-study}(a) highlights their different search paths:
MLE-STAR explores full-catalog masked cross-entropy and loss-weight
ablations, while Self-EvolveRec develops terminal-target curricula
and negative-sampling strategies.
RecHarness uses Bandit-guided arm selection to explore learning
objectives, embedding representations, and coordinated modifications,
followed by feedback-guided refinement.

Figure~\ref{fig:case-study}(b) shows that RecHarness improves validation
HR@10 from 0.5425 to \textbf{0.6275}, exceeding Self-EvolveRec
(0.5988) and MLE-STAR (0.5737).
RecHarness achieves stronger cumulative
improvement through exploration across multiple optimization directions.

\begin{table}[t]
\centering
\small
\setlength{\tabcolsep}{1.8mm}
\begin{tabular}{@{}lccc@{}}
\toprule
Scenario & ADVV & Revenue & Exposure \\
\midrule
Short-video Advertising & +2.084\% & +0.534\% & +0.559\% \\
\bottomrule
\end{tabular}
\caption{Online A/B test in short-video advertising scenario.}
\label{tab:online-ab-main}
\end{table}

\subsection{Online A/B Test}

We deploy RecHarness in a large-scale short-video advertising system whose shared-bottom DNN combines user, item, cross features, and six real-time commerce behavior sequences.
Experts define the objective, validation metric, and edit arms.
After local edits yield no stable gains, the router selects the \texttt{sequence\_\allowbreak encoder\_\allowbreak upgrade} arm.
Guided by Experiment Skill and trial history, RecHarness identifies insufficient relation modeling within raw sequences and introduces intra-sequence self-attention.
Each sequence is compressed to 32 dimensions through an MLP, forming a 192-dimensional representation injected into the ranker.

In a 7-day A/B test on 10\% of production traffic, the candidate improves ADVV (Advertiser Value)~\cite{chai2025longer}, Revenue, and Exposure by \textbf{2.084\%} (\textit{CI: [1.74\%, 2.44\%]}), \textbf{0.534\%} (\textit{CI: [0.34\%, 0.73\%]}), and \textbf{0.559\%} (\textit{CI: [0.44\%, 0.68\%]}), respectively (Table~\ref{tab:online-ab-main}), demonstrating the transfer of automated model improvements to online business gains.

\section{Conclusion}
We introduced RecHarness, a budget-constrained framework that decouples validation-driven Thompson routing over edit directions from LLM-based code mutation. Experiments across two recommendation scenarios and eight models, together with online gains in short-video advertising, demonstrate its effectiveness and practical value.

\section*{Limitations}

RecHarness currently relies on expert-predefined edit arms, and the effectiveness of these arms may change as the incumbent model is updated. Future work could dynamically adjust edit arms using diagnostic evidence and historical trials, and explore routing methods with forgetting mechanisms or context awareness, allowing the search space and trial allocation to adapt to the model state.

\section*{Ethics Statement}

This work studies automated recommender optimization using
Amazon Reviews and KuaiRec, together with a production
advertising experiment. The use of interaction data requires
attention to privacy, dataset licenses, and appropriate
authorization. LLM prompts and released artifacts should
exclude personal identifiers and confidential production
information, and production experiments require appropriate
review and safeguards for affected users.

\bibliography{custom}

\appendix

\section{Related Work}

\subsection{LLM-driven MLE Agents}

Automated machine learning (AutoML) \cite{thornton2013autoWEKA,feurer2015autoML,zheng2023AutoRecSys} development has long been pursued through hyperparameter optimization (HPO) \cite{feurer2019hpo,akiba2019optuna}, and neural architecture search (NAS) \cite{luo2018nao,liu2018darts,ren2021comprehensiveNAS}, but these methods operate within a predefined configuration space. They tune scalar hyperparameters, or search within a fixed architectural template, and therefore struggle with code-level changes that fall outside any enumerable grid, such as redesigning a loss function, switching architectural blocks, or altering a pooling strategy. 
Building on the interleaved reasoning-and-acting paradigm \cite{yao2023react}, LLM-based MLE agents read task descriptions, generate training code, run experiments, and iteratively refine candidate solutions using validation results and execution logs \cite{chan2025mlebench}. A dominant line of work models this as a search problem over candidate programs. AIDE \cite{jiang2025aide} organizes solutions into a tree and adopts a greedy policy that drafts, debugs, or improves the best node. AIRA \cite{toledo2026aira} formalizes such agents as search policies operating over operator sets
These methods are largely gradient-free, compressing execution feedback into scalar validation scores used only to rank and prune branches. More recent work argues that this discards valuable signal: Reasoning-as-Gradient \cite{zhang2026reasoningasgrad} treats error logs, training dynamics, and validation outcomes as textual gradients that guide directed updates, and shows that such directed optimization increasingly surpasses exhaustive tree search as the underlying model's reasoning capability grows.

As one of the most widely deployed applications of ML, recommendation is a natural target for such agents \cite{zheng2023AutoRecSys,wang2022autofea,zhao2021autoloss,lyu2022autofinter}. Its development is inherently iterative: sequential recommendation, ranking/CTR prediction, and watch-time modeling are all shaped by training objectives, optimization strategies, feature interactions, and structural choices that engineers adjust round after round, making it well suited to turning experience-driven, code-level iteration into an executable automated search.

\subsection{Self-Evolving Agentic RecSys}

Since recommendation development is inherently iterative, a growing body of work casts LLM-based agents as automated recommendation engineers that generate, implement, and evaluate model improvements in a closed loop. At industrial scale, Google's dual-loop framework~\cite{wang2026google_recagent} screens hypotheses via cheap proxy metrics in an inner loop and validates candidates against delayed business metrics in an outer loop. GEARS~\cite{yun2026GEARS} encapsulates ranking expertise into reusable agent skills steered by high-level intent. AgentX~\cite{lao2026agentx} runs a four-stage closed loop and distills execution trajectories into semantic-gradient updates that sharpen the agents themselves. EvoRec~\cite{mu2026evorec} co-evolves the model and its methodology by distilling reusable strategies from a persistent experiment memory. NOVA~\cite{liu2026nova} guides architecture evolution with an SGD-inspired architecture gradient aggregating prior edits, diagnostics, and metric feedback. 

However, letting the LLM freely decide the exploration direction is overly divergent: candidates are proposed in an open space without principled trial allocation, yet each edit incurs a costly, noisy training--validation trial, resulting in low success rates and low optimization efficiency \cite{krishnamurthy2024llmexplore,sun2026llmenhancedmab,bouneffouf2026banditsllm}. The key is thus to combine recommendation priors with posterior exploration probabilities as a harness governing where to explore next. 


\section{Algorithm Details}
\label{app:algo}

The main paper defines the budget-constrained optimization objective, the Thompson posterior, the group-normalized routing reward, and the basin-aware jump criterion (Table~\ref{tab:recharness-algorithm-loop}). Here we provide an operational summary of the complete search loop. RecHarness repeatedly routes a limited trial budget over interpretable edit arms, asks the LLM to instantiate arm-conditioned hypotheses as executable mutations, and uses validation-only feedback to update both the incumbent implementation and the routing state. The boxed procedure below lists the per-round steps; each iteration routes the budget to one edit direction, implements it, and promotes only strictly improving candidates.

\begin{table*}[t]
\centering
\begin{minipage}{0.92\textwidth}
\vspace{2pt}
\begin{tcolorbox}[
    enhanced,
    colback=black!3,
    colframe=black!35,
    coltitle=black,
    colbacktitle=black!8,
    boxrule=0.5pt,
    arc=2mm,
    left=3mm,
    right=3mm,
    top=2mm,
    bottom=2mm,
    title=\textbf{RecHarness Bandit Search Loop},
    fonttitle=\small,
    fontupper=\small
]
\begin{enumerate}[leftmargin=*,itemsep=1pt,topsep=1pt]
    \item Evaluate the unmodified implementation $f_0$ and initialize the incumbent score, arm-wise $\mathrm{Beta}(1,1)$ posteriors, Experiment Skill, and validation-score history.
    \item Before round $t$, compute the recent incumbent improvement rate $\widehat{v}_t=(s_t^\star-s_{t-W}^\star)/W$. Jump arms become eligible when $\widehat{v}_t\leq\tau$ and no retuning window is active.
    \item Sample $\tilde{\theta}_a\sim\mathrm{Beta}(\alpha_a,\beta_a)$. A local round selects the top-$G$ eligible local arms. If a jump arm is selected, retain only the jump arm with the largest Thompson sample and execute a single structural candidate.
    \item For every selected arm, use the incumbent or active temporary branch, Experiment Skill, and recent execution feedback to form one compact hypothesis and implement it in an isolated trial workspace.
    \item Train and validate all candidates. For a local round, compute group-normalized routing rewards and promote the highest-scoring valid candidate only when it strictly improves the current incumbent or active branch.
    \item A valid jump candidate creates a temporary branch and saves the pre-jump incumbent as a rollback snapshot. Run up to $R$ local retuning rounds on this branch, with no additional jump allowed during the window.
    \item Apply delayed acceptance after retuning: accept the temporary branch only when its best validation score exceeds the pre-jump incumbent by the required margin; otherwise restore the saved incumbent.
    \item Update the selected arms' posteriors, Experiment Skill, failure memory, score history, and remaining total GPU-time budget.
\end{enumerate}
\end{tcolorbox}
\end{minipage}
\caption{Algorithm Loop of RecHarness.}
\label{tab:recharness-algorithm-loop}
\end{table*}

\subsection{Additional Technical Details}
\label{app:tech}


\paragraph{Promotion and validation isolation.}
All candidates are trained and evaluated in isolated workspaces. Failed, timed-out, unscored, or contract-violating trials consume budget and provide negative routing evidence but cannot be promoted. For a jump, the working implementation may provisionally enter a temporary branch while the pre-jump incumbent is retained as a rollback snapshot; provisional installation is not treated as final jump acceptance. Held-out test data is never read during search.

\paragraph{Jump and retuning state.}
Local arms refine the current basin, whereas jump arms make structural or objective-level changes. Jump eligibility is controlled by the recent improvement rate $\widehat{v}_t\leq\tau$. A jump round contains exactly one jump arm. After a valid jump, RecHarness performs up to $R$ local retuning rounds on the temporary branch and suppresses further jumps during this window. The branch is accepted only after delayed comparison with the saved pre-jump incumbent, and all retuning trials are charged to the same total budget.

\paragraph{Experiment Skill.}
Experiment Skill stores compact lessons from successes, failures, and recent validation trends. It records concrete successful edit patterns, recurring crash or degradation patterns, and textual feedback that helps the LLM avoid repeating known bad patches. It does not replace validation feedback and does not use held-out test metrics during search. Because the skill is updated from routing outcomes, it stays aligned with the dimension-level evidence that drives arm selection, so the LLM's hypothesis for a selected arm reflects what has actually worked on that arm rather than a generic prior over recommender tricks.

\section{Edit-Arm Taxonomy and Safety Constraints}
\label{app:arms}

RecHarness exposes interpretable engineering directions rather than unrestricted code generation. Tables~\ref{tab:amazon-arm-taxonomy} and~\ref{tab:kuairec-arm-taxonomy} summarize the arm families used in the two experimental settings. Composite arms group decisions that are commonly coupled in practice, such as learning rate with batch size and schedule; the LLM still determines the concrete values and the implementation inside the selected direction.

\begin{table*}[t]
\centering
\small
\renewcommand{\arraystretch}{1.2}
\resizebox{\textwidth}{!}{%
\begin{tabular}{
    >{\centering\arraybackslash}p{0.20\textwidth}
    >{\centering\arraybackslash}p{0.36\textwidth}
    >{\centering\arraybackslash}p{0.36\textwidth}}
\toprule
\textbf{Family}
& \textbf{Representative arms}
& \textbf{Intended scope} \\
\midrule
Training configuration
& \texttt{tune\_lr\_batch\_scheduler}, \texttt{tune\_dropout\_wd}
& Coupled optimization, scheduling, and regularization choices. \\
\midrule
Model and context
& \texttt{tune\_embedding\_dim}, \texttt{tune\_num\_layers}, \texttt{tune\_seq\_len}
& Representation capacity, model depth, and history length. \\
\midrule
Jump
& \texttt{change\_architecture}
& Non-local modeling changes that initialize a temporary basin. \\
\bottomrule
\end{tabular}}
\vspace{1pt}\caption{Edit-arm taxonomy for Amazon sequential recommendation.}
\label{tab:amazon-arm-taxonomy}
\end{table*}

\begin{table*}[t]
\centering
\small
\renewcommand{\arraystretch}{1.2}
\resizebox{\textwidth}{!}{%
\begin{tabular}{
    >{\centering\arraybackslash}p{0.20\textwidth}
    >{\centering\arraybackslash}p{0.36\textwidth}
    >{\centering\arraybackslash}p{0.36\textwidth}}
\toprule
\textbf{Family}
& \textbf{Representative arms}
& \textbf{Intended scope} \\
\midrule
Target representation
& \texttt{tune\_vocab\_quantization}
& Dynamic target vocabulary and quantization resolution. \\
\midrule
Objective and decoding
& \texttt{tune\_window\_size}, \texttt{tune\_loss\_balance}
& Soft decoding and ranking--regression objective balance. \\
\midrule
Training configuration
& \texttt{tune\_optimizer\_schedule}, \texttt{tune\_transformer\_capacity}
& Optimization behavior and decoder capacity under the proxy horizon. \\
\midrule
Jump
& \texttt{change\_decoder\_backbone}
& Decoder-family transition that initializes a temporary basin. \\
\bottomrule
\end{tabular}}
\vspace{1pt}\caption{Edit-arm taxonomy for KuaiRec watch-time modeling.}
\label{tab:kuairec-arm-taxonomy}
\end{table*}

The safety policy has three parts. First, mutually interfering arms are not selected in the same local round, so a single round never mixes changes that would confound each other's attribution. Second, a jump round executes exactly one structural arm and blocks further jumps during the subsequent retuning window. Third, every mutation is constrained to approved training and prediction files and must preserve the benchmark entrypoint, evaluation interface, and metric-printing contract. Violations are recorded as failed trials and yield negative routing evidence rather than a repaired candidate.

\section{Prompt Interface and Trial Boundaries}
\label{app:prompt}

The main paper describes the division of labor between routing and LLM reasoning. The following abbreviated prompt illustrates the executor-facing interface used to enforce trial isolation and benchmark compatibility (Table \ref{tab:prompt}). The prompt receives only the routed arm, so the LLM cannot substitute its own direction choice for the bandit decision. The executor never chooses the direction and never sees held-out test data; it only instantiates and runs the routed arm inside the allowed files.

\begin{table*}[t]
\centering
\begin{minipage}{0.98\textwidth}
\vspace{2pt}
\begin{tcolorbox}[
    enhanced,
    colback=black!3,
    colframe=black!35,
    boxrule=0.5pt,
    arc=2mm,
    left=3mm,
    right=3mm,
    top=2mm,
    bottom=2mm,
    title=\textbf{Trial Executor Prompt Skeleton},
    fonttitle=\small,
    fontupper=\small
]
\begin{verbatim}
Input: selected arm, dimension, short hypothesis, Experiment Skill, recent failures, allowed files.

Rules:
- Modify only the allowed training/prediction files.
- Preserve the training entrypoint and metric-printing contract.
- Do not edit benchmark logic, harness state, or parent directories.
- Run the proxy training command before finishing.
- Return changed files, validation metrics, and errors as JSON.
\end{verbatim}
\end{tcolorbox}
\end{minipage}
\vspace{-4pt}
\caption{Abbreviated trial-executor prompt used to bound a single trial.}
\label{tab:prompt}
\end{table*}

\section{Additional Experimental Settings}
\label{app:setup}

\subsection{Tasks and Datasets}
\label{app:data}
We evaluate RecHarness on two recommendation settings:
next-item prediction on Amazon Reviews and watch-time prediction on KuaiRec. Tables
\ref{tab:amazon-dataset-stats} and \ref{tab:kuairec-dataset-stats}
summarize the dataset statistics after preprocessing.

For sequential recommendation, we use four Amazon Reviews 5-core datasets: Movies, Scientific, Electronics, and CDs. The 5-core setting ensures that each retained user and item has at least five interactions. For watch-time and ranking prediction, we use KuaiRec, whose dense user--video interaction data supports watch-time, watch-ratio, and ranking-related objectives.

\begin{table}[t]
\centering
\setlength{\tabcolsep}{1.0mm}
\begin{tabular}{cccc}
\toprule
Dataset & \#Users & \#Items & \#Interactions \\
\midrule
Movies & 11,947 & 17,490 & 144,071 \\
Scientific & 23,627 & 25,764 & 266,164 \\
Electronics & 27,601 & 31,533 & 292,308 \\
CDs & 18,481 & 30,951 & 284,695 \\
\bottomrule
\end{tabular}
\caption{Statistics of the Amazon Reviews datasets.}
\label{tab:amazon-dataset-stats}
\end{table}

\begin{table}[t]
\centering
\small
\setlength{\tabcolsep}{1.0mm}
\resizebox{\columnwidth}{!}{
\begin{tabular}{@{}ccccc@{}}
\toprule
Split & Interactions & Users & Items & Avg. Inter./User \\
\midrule
Train & 12,530,806 & 7,176 & 10,728 & $\sim$ 1,746 \\
Test  & 4,676,570  & 1,411 & 3,327  & $\sim$ 3,314 \\
\bottomrule
\end{tabular}}
\caption{Statistics of the KuaiRec dataset.}
\label{tab:kuairec-dataset-stats}
\end{table}

\subsection{Baselines} \label{app:base}
We compare RecHarness with the initial recommender
implementation (\textbf{Base}) and two LLM-based
optimization frameworks, covering general-purpose
machine learning engineering and recommendation-specific
model evolution.

\begin{itemize}
    \item \textbf{MLE-STAR \cite{nam2025mlestar}}.
    MLE-STAR is a machine learning engineering agent
    that combines web-based model search with targeted
    code refinement. It retrieves task-relevant modeling
    approaches to construct an initial solution, then
    uses code-block ablation studies to identify
    influential components and guide subsequent
    refinement. Its framework also includes an
    ensembling stage.

    \item \textbf{Self-EvolveRec \cite{kim2026self}}.
    Self-EvolveRec is a recommendation-specific
    code-evolution framework that guides model
    improvement through directional feedback. It
    combines a User Simulator for qualitative
    critiques with a Model Diagnosis Tool for
    quantitative verification of model behavior.
    Its diagnosis-tool--model co-evolution strategy
    adapts the diagnostic tools as the recommender
    architecture evolves.
\end{itemize}

Together, these baselines evaluate RecHarness against
both general-purpose targeted refinement and
recommendation-specific directional feedback.

\subsection{Model Templates}
\label{app:templates}

For Amazon sequential recommendation, RecHarness optimizes five implementation templates:
\begin{itemize}[leftmargin=*]
    \item \textbf{GRU4Rec}: a recurrent sequential recommender that updates a hidden state step by step, representing short-term sequential dependency modeling.
    \item \textbf{BERT4Rec}: a bidirectional Transformer recommender whose masked-item objective allows each position to use both left and right context during training.
    \item \textbf{NextItNet}: a convolutional sequential recommender where dilated convolutions enlarge the receptive field while preserving parallelizable sequence modeling.
    \item \textbf{SASRec}: a causal self-attention recommender that adaptively attends to relevant historical items.
    \item \textbf{HSTU}: a stronger hierarchical sequential template used to test whether RecHarness can improve a more complex sequential model template under the same local evaluation protocol.
\end{itemize}

For KuaiRec, RecHarness optimizes three watch-time or response-modeling templates:
\begin{itemize}[leftmargin=*]
    \item \textbf{D2Q}: converts continuous watching feedback into discrete target bins, testing RecHarness under a classification-style target formulation.
    \item \textbf{TPM}: directly optimizes watch-time and watch-ratio feedback, testing RecHarness on continuous user-video response modeling.
    \item \textbf{GR}: uses a generative regression formulation, testing whether RecHarness can improve models that jointly balance ranking quality and numerical prediction error.
\end{itemize}

The five Amazon templates span recurrent, bidirectional Transformer, convolutional, causal self-attention, and hierarchical formulations, while the three KuaiRec templates span classification-style discretization, continuous response modeling, and generative regression. This coverage matters for evaluating a harness rather than a single architecture: an edit arm that helps one template may be neutral or harmful for another, so the harness must allocate trials per template instead of committing to one global recipe.

\subsection{Metrics and Protocol}
\label{app:metrics}
For sequential recommendation, we use leave-last-out evaluation: the last item is used for testing, the second-to-last for validation, and earlier interactions for training. Each evaluation set contains one positive item and 99 sampled negatives. During Amazon search, validation HR@10 is the sole metric used for routing rewards, posterior updates, candidate promotion, jump triggering, and delayed jump acceptance. After search terminates, we evaluate the selected implementation once on the held-out test set and report Hit Ratio (HR@N) and Normalized Discounted Cumulative Gain (NDCG@N), with $N$ set to 10 and 20. For a compact Amazon summary, we use \textit{Avg. HR@10 over datasets}:
\begin{equation}
    \mathrm{AvgHR@10}(m)=\frac{1}{|\mathcal{D}_{\mathrm{A}}|}
    \sum_{d\in\mathcal{D}_{\mathrm{A}}}\mathrm{HR@10}(m,d),
\end{equation}
where $m$ is a method or model variant and $\mathcal{D}_{\mathrm{A}}$ is the Amazon dataset set. Per-dataset metrics remain the primary evidence.

For KuaiRec, let $y_i^{\mathrm{WT}}$ and $\widehat{y}_i^{\mathrm{WT}}$ denote the observed and predicted watch time, measured in seconds, for interaction $i$. Let $d_i$ be the video duration and define the predicted watch ratio as $\widehat{y}_i^{\mathrm{WR}}=\widehat{y}_i^{\mathrm{WT}}/\max(d_i,\varepsilon)$; the observed watch ratio is denoted $y_i^{\mathrm{WR}}$. WT and WR denote watch-time and watch-ratio. Lower MAE and higher XAUC are better \cite{ma2024generative}. We report four metrics:

For a target vector $y$ and prediction vector $\widehat{y}$, XAUC is the fraction of unordered interaction pairs whose relative order is correctly predicted:
\begin{equation}
    \scalebox{0.95}{$\displaystyle
    \mathrm{XAUC}(y,\widehat{y})
    =\frac{1}{\binom{n}{2}}\sum_{i<j}
    \mathbb{I}\big[(y_i-y_j)(\widehat{y}_i-\widehat{y}_j)>0\big]$},
\end{equation}

\begin{itemize}[leftmargin=*]
    \item \textbf{WT-XAUC} applies this definition to $y^{\mathrm{WT}}$ and $\widehat{y}^{\mathrm{WT}}$. It evaluates whether videos with longer realized watch time receive larger predictions; higher is better.
    \item \textbf{WT-MAE} measures absolute watch-time prediction error in seconds:
    \begin{equation}
        \operatorname{WT\text{-}MAE}
        =\frac{1}{n}\sum_{i=1}^{n}
        \left|\widehat{y}_i^{\mathrm{WT}}-y_i^{\mathrm{WT}}\right|.
    \end{equation}
    Lower WT-MAE indicates more accurate numerical watch-time estimates.
    \item \textbf{WR-XAUC} applies the same pairwise agreement measure to observed and predicted watch ratios, $y^{\mathrm{WR}}$ and $\widehat{y}^{\mathrm{WR}}$. It evaluates duration-normalized ranking quality and is higher-is-better.
    \item \textbf{WR-MAE} measures the mean absolute error of the duration-normalized watch ratio:
    \begin{equation}
        \operatorname{WR\text{-}MAE}
        =\frac{1}{n}\sum_{i=1}^{n}
        \left|\widehat{y}_i^{\mathrm{WR}}-y_i^{\mathrm{WR}}\right|,
    \end{equation}
    and is dimensionless and lower-is-better.
\end{itemize}

WT metrics emphasize absolute engagement duration, whereas WR metrics normalize engagement by video length. XAUC evaluates ranking consistency, while MAE evaluates pointwise numerical accuracy; reporting both prevents improvements in ordering quality from obscuring calibration error, or vice versa. A model can rank videos correctly while systematically over-predicting duration, or match durations closely while misordering near-tied pairs, so the four metrics are complementary rather than redundant.

\subsection{Reproducibility and Hyperparameters}
\label{app:hyper}

Table~\ref{tab:cold-start-configs} summarizes the core cold-start settings. These values define the initial templates; a trial changes them only when its selected edit arm targets the corresponding dimension.

\begin{table*}[t]
\centering
\small
\resizebox{\textwidth}{!}{
\begin{tabular}{lllll}
\toprule
Task & Template & Core architecture & Dropout & Optimization \\
\midrule
 & GRU4Rec & hidden 128; 1 GRU layer & 0.1 & Adam, LR $10^{-3}$ \\
 & BERT4Rec & hidden 64; 2 blocks; 2 heads & 0.2 & Adam, LR $10^{-3}$ \\
Amazon & NextItNet & hidden 64; 8 blocks; kernel 3 & 0.2 & Adam, LR $10^{-3}$ \\
 & SASRec & hidden 64; 2 blocks; 1 head & 0.5 & Adam, LR $10^{-3}$ \\
 & HSTU & hidden 64; 2 blocks; 2 heads & 0.3 & AdamW, LR $10^{-3}$, WD $10^{-6}$ \\
 \midrule
 & D2Q & emb. 32; hidden 256; 32 groups & 0.2 & Adagrad, LR $10^{-3}$ \\
KuaiRec & TPM & emb. 16; hidden 128; 32 leaves & 0.2 & Adam, LR $10^{-3}$ \\
 & GR & hidden 128; 3 layers; 8 heads & 0.1 & Adam, LR $5\!\times\!10^{-4}$ \\
\bottomrule
\end{tabular}}
\caption{Core hyperparameters of the cold-start templates.}
\label{tab:cold-start-configs}
\end{table*}

The shared training and search settings are reported in Table~\ref{tab:training-search-configs}. Amazon templates use sequence length 128 and 200 gradient steps per epoch. Validation is performed every five epochs, with early stopping based on the template's validation patience. The KuaiRec two-epoch setting is the fixed proxy horizon used during search.

\begin{table}[t]
\centering
\small
\setlength{\tabcolsep}{1.2mm}
\begin{tabular}{@{}>{\raggedright\arraybackslash}p{0.42\columnwidth}>{\raggedright\arraybackslash}p{0.50\columnwidth}@{}}
\toprule
Setting & Value \\
\midrule
LLM & GPT-6.1-Sol \\
Candidates per local round & $G=4$ parallel candidates \\
Candidates per jump round & 1 structural candidate \\
Thompson prior & $\mathrm{Beta}(1,1)$ for each arm \\
Jump trigger & $\widehat{v}_t\leq\tau$, with $\tau=0.03$ \\
Post-jump retuning & $R=4$ local rounds; no new jump during retuning \\
Jump acceptance & Delayed comparison with the pre-jump incumbent \\
Per-model search budget & 43,200 s total GPU-time \\
\bottomrule
\end{tabular}
\caption{Shared training and search hyperparameters.}
\label{tab:training-search-configs}
\end{table}

A jump round evaluates one structural candidate rather than four local candidates. The jump gate opens when the recent-round improvement rate satisfies $\widehat{v}_t\leq0.03$ and no retuning window is active. Each model receives 43,200 s of total GPU-time: for a parallel round, the runtimes of all concurrently evaluated candidates are summed rather than counted as a single wall-clock duration. All routing variants use the same data split, training command, validation protocol, promotion rule, and total GPU-time budget.

\section{Additional Experiments}

\begin{table*}[t]
\centering
\small
\setlength{\tabcolsep}{2.4mm}
\begin{tabular}{@{}lcccc@{}}
\toprule
Method & Improving / All & Ratio & Avg Gain & Max Gain \\
\midrule
RecHarness & 23 / 48 & 47.92\% & 5.06\% & 24.00\% \\
TR w/ Random & 11 / 49 & 22.45\% & 4.94\% & 16.63\% \\
TR w/ LLM & 10 / 46 & 21.74\% & 5.04\% & 15.11\% \\
w/o Bandit & 20 / 48 & 41.67\% & 4.05\% & 10.16\% \\
\bottomrule
\end{tabular}
\caption{Trial-level gains relative to the round-start best validation score.}
\label{tab:rq3-trial-gains}
\end{table*}

We measure how often non-baseline trials improve the round-start best validation score, and how large those improvements are, as shown in Table~\ref{tab:rq3-trial-gains}. The results show two effects of the RecHarness routing design. First, RecHarness finds improvements more accurately than the two routing ablations: 47.92\% of its non-baseline trials refresh the round-start best score, compared with only 22.45\% for \textit{TR w/ Random } replacement and 21.74\% for \textit{TR w/ LLM}. This indicates that RecHarness does more than record past outcomes: it uses accumulated validation feedback to allocate future trials toward edit directions that are empirically more likely to improve the current best model. Second, RecHarness finds higher-quality improvements than the \textit{w/o Bandit}. Although \textit{w/o Bandit} still improves 41.67\% of trials, its average and maximum gains are both lower than those of RecHarness, especially in maximum gain (10.16\% vs. 24.00\%). This suggests that unrestricted LLM search can still discover useful directions, but it's less likely than bandit routing to allocate trials to high-upside directions.

Thus, RecHarness improves mainly by allocating limited trials more effectively. Together, the LLM proposes hypotheses within selected directions, while bandit routing makes limited trials more likely to produce large gains.

\section{Extended Results}
\label{app:extended}

\paragraph{Representative edit patterns.}

Across the eight optimized templates, the most useful edits are usually simple and recommender-specific rather than large rewrites. For sequential recommendation, high-value directions concentrate on regularization, optimization, pooling, and moderate capacity changes. SASRec benefits from coupled dropout--weight-decay tuning, batch-size adjustment, and attention-head tuning. HSTU benefits from loss alignment, weight-decay tuning, hybrid last/mean pooling, and lightweight temporal or recency features. BERT4Rec and NextItNet show similar sensitivity to weight decay, dropout, scheduler choice, and depth.

For KuaiRec watch-time modeling, useful edits are more closely tied to target formulation. D2Q and TPM benefit from optimizer-schedule changes, duration-aware modeling, loss rebalancing, and quantization or leaf-resolution changes. GR starts from a stronger baseline, so its gains are smaller and mostly come from low-risk changes such as cosine scheduling, curriculum adjustment, and regression-loss weighting.

Two observations follow. First, the winning edits are rarely exotic: they are the same controls a recommender engineer would sweep first, which suggests the value of the harness lies in allocation rather than in proposing novel architectures. Second, the useful direction depends on the template's starting point. Stronger baselines such as HSTU and GR leave less room for structural change and are improved mainly by regularization and objective alignment, whereas weaker templates still gain from capacity and depth adjustments. These observations support the use of structured edit arms: the space is broad enough to cover meaningful recommender-engineering choices, but constrained enough to avoid spending most trials on arbitrary rewrites.

\section{Case Studies}
\label{app:cases}

This section inspects concrete search traces to illustrate how RecHarness allocates exploration under a limited trial budget, how an occasional random success should be interpreted, and how a basin jump can relocate subsequent local search. These traces are intended as mechanistic examples rather than independent statistical evidence.

\subsection{Exploration Pressure after the First Improvement}

The first case focuses on post-success search behavior in the SASRec controlled ablation, comparing full Thompson routing with TR w/ LLM. We examine how each method allocates trials after the first validation improvement. Specifically, we count how often post-success trials reuse the first successful arm, reuse any previously successful arm, explore an arm that had not appeared before the first improvement, and further improve the round-start best. Table~\ref{tab:case-exploration-pressure} reports the resulting statistics.

\begin{table*}[t]
\centering
\small
\setlength{\tabcolsep}{1.6mm}
\begin{tabular}{@{}lccccc@{}}
\toprule
Method & First Winner & Prior Winner & New Arm & Improve Ratio & Gain \\
\midrule
Full Thompson & 25.69\% & 49.31\% & 52.08\% & 49.31\% & 0.1062 \\
TR w/ LLM & 10.26\% & 15.38\% & 81.41\% & 15.38\% & 0.0074 \\
\bottomrule
\end{tabular}
\caption{Post-success exploration behavior in the SASRec controlled ablation. \textit{First Winner} is the share of post-success trials that reuse the first improving arm, \textit{Prior Winner} the share reusing any previously successful arm, \textit{New Arm} the share exploring an unseen arm, and \textit{Improve Ratio} the share that further improve the round-start best. \textit{Gain} is the total validation HR@10 improvement over the run.}
\label{tab:case-exploration-pressure}
\end{table*}

Full Thompson does not simply repeat the first successful arm: only 25.69\% of post-success trials reuse it, while 52.08\% explore new arms. It also preserves exploitation pressure, with 49.31\% of post-success trials reusing some previously successful arm and 49.31\% further improving the round-start best. In contrast, TR w/ LLM explores more broadly after success, with 81.41\% new-arm trials, but only 15.38\% of post-success trials improve the round-start best. This suggests that posterior routing maintains calibrated exploration in a structured edit-dimension space, rather than drifting toward either pure repetition or pure novelty.

\subsection{Random Routing Hits a Regularization Gain}

In the TR w/ Random ablation, we observe one high-reward random-arm run. It is the best of the three random-arm runs: validation best HR@10 improves from 0.5050 to 0.6075, and the final held-out test aggregate HR@10 is 0.5381. However, the average final validation best over the three random-arm runs is still 0.5742, below the full RecHarness average of 0.6371. Table~\ref{tab:case-random-trace} summarizes the winning trajectory.

\begin{table}[t]
\centering
\footnotesize
\setlength{\tabcolsep}{0.8mm}
\begin{tabular}{@{}l>{\ttfamily\scriptsize}l>{\footnotesize}r>{\footnotesize}r>{\footnotesize}r@{}}
\toprule
Stage & \normalfont\footnotesize Winning arm & \footnotesize HR@10 & \footnotesize$\Delta$base & \footnotesize$\Delta$prev \\
\midrule
Base & -- & 0.5050 & -- & -- \\
R1 & tune\_num\_heads & 0.5112 & +0.0062 & +0.0062 \\
R2 & tune\_dropout\_wd & 0.5962 & +0.0912 & +0.0850 \\
R3 & tune\_lr\_batch\_scheduler & 0.6013 & +0.0963 & +0.0051 \\
R4 & tune\_weight\_decay & 0.6075 & +0.1025 & +0.0062 \\
\bottomrule
\end{tabular}
\caption{Winning trajectory of the highest-scoring random-arm SASRec run. $\Delta$base and $\Delta$prev are relative to the baseline and the previous round.}
\label{tab:case-random-trace}
\end{table}

The final mutation stack contains \texttt{NUM\_HEADS=2}, \texttt{DROPOUT\_RATE=0.3}, \texttt{WEIGHT\_DECAY=5e-4}, \texttt{BATCH\_SIZE=256}, and step learning-rate decay. In this run, the high reward of random routing mainly comes from hitting \texttt{tune\_dropout\_wd}; capacity and optimization-schedule changes contribute smaller follow-up gains. The trace supports a limited claim: random routing can occasionally hit a strong direction under a small budget, but it does not imply that random routing is stable, as the three-run average in the main paper shows.

\subsection{Search Relocation after Basin Jump}

The final illustrative case comes from HSTU. The run improves from baseline 0.5325 to 0.5700 over the first four rounds, with winning directions including \texttt{tune\_num\_layers}, \texttt{tune\_dropout\_wd}, and \texttt{change\_pooling}. At iteration 5, the router selects the single jump arm \texttt{change\_loss\_function}. Its immediate validation score is 0.5162, below the pre-jump incumbent of 0.5700. RecHarness therefore retains 0.5700 as the rollback reference while using the jump candidate as a temporary branch for subsequent local retuning. Table~\ref{tab:case-basin-jump} shows the key trajectory.

\begin{table}[t]
\centering
\footnotesize
\setlength{\tabcolsep}{1.0mm}
\begin{tabular}{@{}cl>{\ttfamily\scriptsize}l>{\footnotesize}r>{\footnotesize}r@{}}
\toprule
Iter. & \footnotesize Event & \normalfont\footnotesize Arm & \footnotesize Branch & \footnotesize Incumbent \\
\midrule
1 & base & -- & 0.5325 & 0.5325 \\
2 & local & tune\_num\_layers & 0.5575 & 0.5575 \\
3 & local & tune\_dropout\_wd & 0.5675 & 0.5675 \\
4 & local & change\_pooling & 0.5700 & 0.5700 \\
5 & jump & change\_loss\_function & 0.5162 & 0.5700 \\
6 & retune & tune\_weight\_decay & 0.6138 & 0.6138 \\
7 & retune & change\_pooling & 0.6262 & 0.6262 \\
\bottomrule
\end{tabular}
\caption{HSTU basin-jump trace. \textit{Branch} is the validation HR@10 of the candidate or temporary branch and \textit{Incumbent} the accepted best; the incumbent is unchanged when a candidate is rejected.}
\label{tab:case-basin-jump}
\end{table}

The conservative interpretation is that the basin jump acts as a relocation mechanism. The direct jump score is not sufficient for acceptance, but retuning the temporary branch discovers \texttt{tune\_weight\_decay} and \texttt{change\_pooling}, raising validation HR@10 to 0.6262. The branch is accepted only after this delayed comparison exceeds the saved pre-jump incumbent. We use this trace only as a mechanistic example, not as statistical evidence that basin jumps are always beneficial.

\section{Representative Search Log}
\label{app:log}

The following condensed log shows one complete RecHarness run for the Amazon
NextItNet template. NextItNet is trained independently on Movies and TV,
Industrial and Scientific, Electronics, and CDs and Vinyl, while sharing one
model configuration across the four datasets. Search decisions use only
validation AvgHR@10. The displayed arm names follow the compact edit-arm
taxonomy introduced above.

\medskip\noindent\rule{\linewidth}{0.5pt}\par
\noindent{\small\textbf{Condensed RecHarness Log: Amazon NextItNet}}\par\medskip
\noindent
{\ttfamily\scriptsize\raggedright\setlength{\parskip}{2pt}\setlength{\parindent}{0pt}
\textbf{[Setup]} model = NextItNet-per-dataset

\textbf{[Setup]} datasets = Movies and TV, Industrial and Scientific,
Electronics, CDs and Vinyl

\textbf{[Setup]} routing metric = validation AvgHR@10

\textbf{[Setup]} candidates per local round = 4

\textbf{[Setup]} jump candidates per jump round = 1

\textbf{[Setup]} total GPU-time budget = 43,200 s

\medskip
\hrule
\medskip

\textbf{[Cold start] No edit arm}

validation AvgHR@10 = 0.5425

decision = establish initial incumbent

cumulative GPU-time = 407 s
remaining budget = 42,793 s

\medskip
\hrule
\medskip

\textbf{[Local promotion 1] Model and context}

candidate validation AvgHR@10 = 0.5587

decision = promoted

cumulative GPU-time = 3,881 s
remaining budget = 39,319 s

\medskip
\hrule
\medskip

\textbf{[Local promotion 2] Model and context}

candidate validation AvgHR@10 = 0.5625

decision = promoted

cumulative GPU-time = 9,063 s
remaining budget = 34,137 s

\medskip
\hrule
\medskip

\textbf{[Local plateau]}

Three consecutive candidate groups produced no incumbent improvement.

best candidate validation AvgHR@10 = 0.5625

incumbent validation AvgHR@10 = 0.5625

decision = reject all candidates; open the jump gate

cumulative GPU-time = 14,295 s
remaining budget = 28,905 s

\medskip
\hrule
\medskip

\textbf{[Temporary jump 1] Jump}

selected arm = \texttt{change\_loss\_function}

temporary-branch AvgHR@10 = 0.5525

saved incumbent AvgHR@10 = 0.5625

decision = retain the candidate as a temporary branch for local retuning

cumulative GPU-time = 14,668 s
remaining budget = 28,532 s

\medskip
\hrule
\medskip

\textbf{[Post-jump retuning and delayed acceptance]}

number of local retuning groups = 4

best retuned-branch AvgHR@10 = 0.5737

pre-jump incumbent AvgHR@10 = 0.5625

delayed-acceptance improvement = +0.0112

decision = accept the retuned temporary branch

cumulative GPU-time = 21,023 s
remaining budget = 22,177 s

\medskip
\hrule
\medskip
\par}
{\ttfamily\scriptsize\raggedright\setlength{\parskip}{2pt}\setlength{\parindent}{0pt}
\textbf{[Second local plateau]}

Three consecutive candidate groups produced no incumbent improvement.

best candidate validation AvgHR@10 = 0.5737

incumbent validation AvgHR@10 = 0.5737

decision = reject all candidates; open the jump gate

cumulative GPU-time = 22,645 s
remaining budget = 20,555 s

\medskip
\hrule
\medskip

\textbf{[Temporary jump 2] Jump}

selected arm = \texttt{change\_architecture}

temporary-branch AvgHR@10 = 0.5625

saved incumbent AvgHR@10 = 0.5737

decision = retain the candidate as a temporary branch for local retuning

cumulative GPU-time = 22,976 s
remaining budget = 20,224 s

\medskip
\hrule
\medskip

\textbf{[Post-jump retuning and rollback]}

number of local retuning groups = 4

best retuned-branch AvgHR@10 = 0.5650

pre-jump incumbent AvgHR@10 = 0.5737

delayed-acceptance improvement = -0.0087

decision = reject the retuned temporary branch; restore the saved incumbent

cumulative GPU-time = 28,631 s
remaining budget = 14,569 s

\medskip
\hrule
\medskip

\textbf{[Remaining rejected groups]}

best candidate validation AvgHR@10 = 0.5737

incumbent validation AvgHR@10 = 0.5737

decision = reject all candidates; retain the incumbent

cumulative GPU-time = 43,165 s
remaining budget = 35 s

\medskip
\hrule
\medskip

\textbf{[Budget termination]}

required budget for another candidate group $>$ 35 s

decision = terminate search without launching another group

final validation AvgHR@10 = 0.5737

total search consumption = 43,165 / 43,200 s (99.9\%)

\medskip
\hrule
\medskip

\textbf{[Final held-out test evaluation]}

The selected implementation is evaluated once after search termination.
The held-out test metrics are not used for routing, promotion, jump
triggering, or delayed acceptance.

\begin{center}
\renewcommand{\arraystretch}{1.15}
\resizebox{\linewidth}{!}{
\begin{tabular}{lrrrr}
\toprule
Dataset
& HR@10
& HR@20
& NDCG@10
& NDCG@20 \\
\midrule
Movies and TV
& 0.5203
& 0.6471
& 0.3383
& 0.3703 \\
Industrial and Scientific
& 0.5021
& 0.6454
& 0.3062
& 0.3424 \\
Electronics
& 0.4237
& 0.5674
& 0.2591
& 0.2953 \\
CDs and Vinyl
& 0.5498
& 0.6933
& 0.3470
& 0.3833 \\
\midrule
Macro average
& 0.4990
& 0.6383
& 0.3126
& 0.3478 \\
\bottomrule
\end{tabular}}
\end{center}

\textbf{[End]} search budget accounting excludes the final held-out test run.
\par}
\par\medskip\noindent\rule{\linewidth}{0.5pt}\par

\end{document}